\documentclass[usenatbib]{mn2e}
\usepackage{times}
\usepackage{graphicx}
\begin{document}
\title[Lithium-rich giants in SDSS]{Lithium-rich field giants in the Sloan Digital Sky Survey}
 
\bigskip
\author[S. L. Martell \& M. D. Shetrone]{S. L. Martell$^{1}$\thanks{Email: smartell@aao.gov.au (SLM); shetrone@as.utexas.edu (MDS)} and M. D. Shetrone$^{2}$ \\
$^1$Australian Astronomical Observatory, North Ryde NSW 2122, Australia\\
$^2$McDonald Observatory, University of Texas Austin, McDonald Observatory, Texas 79734, USA}

\date{Received 25 October 2012/ Accepted 19 December 2012}

\maketitle

\begin{abstract}
  We present a search for post-main-sequence field stars in the Galaxy with atypically large lithium abundances. Using moderate-resolution spectra taken as part of the Sloan Digital Sky Survey, along with high-resolution followup spectroscopy from the Hobby-Eberly Telescope, we identify $23$ post-turnoff stars with ${\rm log} \epsilon$(Li) greater than 1.95, including 14 with ${\rm log} \epsilon{\rm (Li)} \geq 2.3$ and 8 with ${\rm log} \epsilon{\rm (Li)} \geq 3.0$, well above the low level expected for evolved stars. Comparison with theoretical isochrones indicates that some of our Li-rich stars are affiliated with the upper red giant branch, the asymptotic giant branch and the red clump  rather than the RGB bump, which is a challenge to existing models of Li production in evolved stars.
\end{abstract} 


\section{Introduction}
The abundance of lithium is extremely sensitive to the ambient conditions inside stars, since its cross section to proton capture is quite large at temperatures above $2.6 \times 10^{6} {\rm K}$. Li is readily destroyed in the interiors of low-mass main-sequence stars, and the Li in the convective envelope is progressively depleted as envelope material is cycled past the innermost, hottest region of the convective zone. Surface Li abundance is further depleted after stars leave the main sequence and undergo first dredge-up \citep{I68}, in which the convective envelope deepens, mixing fusion-processed material from the stellar interior into the envelope and exposing envelope material to even higher temperatures. 

Because of this, it was quite surprising when \citet{WS82} discovered a Li-rich K giant serendipitously while investigating barium-rich field stars. More systematic searches for Li-rich stars (e.g., \citealt{B89}; \citealt{P86}; \citealt{PS88}; \citealt{PSK00}) found that they are rare in the Solar neighborhood and extremely unusual in globular clusters. The study of \citet{CB00} found that nearby Solar-metallicity Li-rich stars are either low-mass stars ($M\simeq 1 M_{\sun}$) near the ``bump'' in the red giant branch (RGB) luminosity function or intermediate-mass stars ($M\simeq 3 M_{\sun}$) low on the asymptotic giant branch (AGB). In both cases, the hydrogen-burning shell in the stars in question has recently encountered, and destroyed, the discontinuity in mean molecular weight (the ``$\mu$-barrier'') left behind at the end of first dredge-up. In low-mass red giants, the destruction of the $\mu$-barrier causes a brief loop in stellar evolution, observed as the RGB bump, and permits deep mixing, a nonconvective process that steadily cycles material between the photosphere and the radiative zone, causing surface abundances of carbon and nitrogen to evolve constantly as stars ascend the RGB (see \citealt{MSB08c} for a more thorough explanation of deep mixing). In higher-mass stars, the destruction of the $\mu$-barrier occurs at {\bf second} dredge-up, low on the asymptotic giant branch. The fact that the Li-rich stars identified by \citet{CB00} were located near the destruction of the $\mu$-barrier led those authors to speculate that the onset of deep mixing produces a brief episode of Li production via the Cameron-Fowler mechanism \citep{CF71}. In this process, $^{7}{\rm Be}$ produced by the proton-proton chain is transported to a cooler zone in the star, where it captures an electron and decays into $^{7}{\rm Li}$, which is stable against proton capture if its surroundings are cool enough. A phase in which the deep-mixing mass transport speed was briefly high enough to allow this process would cause surface Li abundances to rise sharply and then decay.

However, the existence of Li-rich stars in other regions of the Hertzsprung-Russell diagram challenges this model: \citet{K99} found one bright red giant well above the RGB bump in the globular cluster M3, where deep mixing is well known to operate (\citealt{S02}; \citealt{AC11}). More recently, six independent studies in the past three years have reported a total of 47 new Li-rich post-main-sequence stars. These new Li-rich stars fill a broad range of parameter space: low-mass ($M\le 2M_{\sun}$) RGB stars in the thick disk with $-1\le{\rm [Fe/H]}\le 0$ \citep{MV11}; an intermediate-mass ($M\simeq 4M_{\sun}$), Solar-metallicity star in the thin disk \citep{MV11}; low-mass RGB stars with ${\rm [Fe/H]} \simeq -0.9$ in the Galactic bulge (\citealt{GZ09}; \citealt{LU12}); and low-mass RGB stars in Local Group dwarf galaxies, two of which have $\rm{[Fe/H]} \simeq -2.7$ \citep{KF12}.

Two of the recent studies (\citealt{KR11} and \citealt{RF11}) take the approach of selecting a well-defined subset from an existing catalog ({\it Hipparcos} and RAVE, respectively) for high-resolution spectroscopic followup and abundance analysis, a procedure that we also adopt. In this paper, we present a search for Li-rich field stars in Sloan Digital Sky Survey (SDSS) spectroscopy. By surveying stars across a wide range of metallicity and surface gravity, we are able to take a broad view of the question of Li-rich post-main-sequence stars. This paper describes the 23 previously unknown Li-rich stars we identified from this original data set, and explores whether the various proposed mechanisms for Li production in evolved stars are sufficient to explain this group of Li-rich stars.

\begin{figure}
\includegraphics[width=84mm]{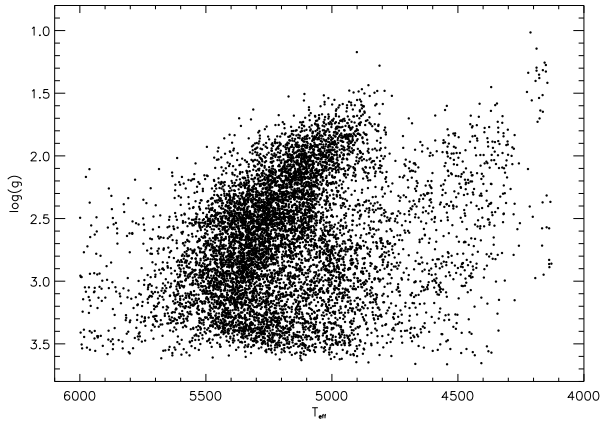}
\caption{Distribution of ${\rm T}_{\rm eff}$ and ${\rm log(g)}$ for
the 8535 stars in our initial data set.}
\end{figure}

\section{The observational data set}
In order to explore the physical conditions that produce large amounts of photospheric lithium in some evolved stars, we selected a sample of post-main-sequence stars from the SDSS Data Release 7. The 8535 stars in our initial data set were required to have high-quality spectra and relatively low errors on stellar parameters determined by the automated SSPP pipeline \citep{LBS08a}, and were allowed to cover a broad range of metallicity, effective temperature and surface gravity. 

\subsection{SDSS spectroscopy}
The imaging component of the original Sloan Digital Sky Survey and its extensions \citep{FI96,GC98,Y00,PM03,GS06,SL02,DR1,DR2,DR3,DR4,DR5,DR6,DR7,DR8} includes $ugriz$ photometry for several hundred million stars. SEGUE \citep{SEGUE}, one of three sub-surveys that together formed SDSS-II, extended that imaging footprint by approximately 3500 deg${\rm ^{2}}$, and also obtained $R \simeq 2000$ spectroscopy for approximately $240,000$ stars over a wavelength range of $3800-9200\hbox{\AA}$. Additional spectroscopy with similar characteristics has been taken as part of the SDSS Legacy project \citep{DR7} and in preparation for the SDSS-III Multi-object APO Radial Velocity Exoplanet Large-area Survey (MARVELS) project \citep{MARVELS}. To facilitate the development and calibration of the SEGUE Stellar Parameter Pipeline (SSPP; \citealt{LBS08a}; \citealt{LBS08b}; \citealt{AP08}), SEGUE included spectra for a collection of Galactic globular and open clusters. As described in \citet{LBS08a}, SSPP produces estimates of ${\rm T_{eff}}$, ${\rm log(g)}$, ${\rm [Fe/H]}$, and radial velocity, along with the equivalent widths and/or line indices for $85$ atomic and molecular absorption lines, by processing the calibrated spectra generated by the standard SDSS spectroscopic reduction pipeline \citep{SL02}. Additional validations and details of recent updates to SSPP are discussed in \citet{SL11}. 

\subsection{Target selection}
We selected our sample from the seventh SDSS data release (DR7), using the online Catalog Archive Server (CAS)\footnote{http://skyservice.pha.jhu.edu/CasJobs/login.aspx} to identify stars with a mean signal-to-noise ratio per pixel (SNR) of at least $40$, ${\rm T}_{\rm eff} \leq 6000 {\rm K}$, reasonable errors on ${\rm [Fe/H]}$ metallicity ($\sigma_{\rm [Fe/H]} \leq 0.3$) and surface gravity ($\sigma_{\rm log(g)} \leq 0.5$), and surface gravity no higher than 1.4 dex greater than the surface gravity at the ``bump'' in the RGB luminosity function. Previous studies have found that stars with unexpectedly high lithium abundances tend to be near the RGB bump, so we designed this gravity selection to include stars at evolutionary phases from shortly before the RGB bump to the tip of the RGB. Empirically, the luminosity of the RGB bump is a function of metallicity: \citet{FP90} report ${\rm [Fe/H]}$ versus $M_{V}^{\rm bump}$ for a selection of Galactic globular clusters, and we converted those $M_{V}$ values to ${\rm log(g)}$ using metallicity-matched 12 Gyr Yale-Yonsei isochrones \citep{D04}, then fit a least-squares line to derive ${\rm log(g)}_{\rm bump} = 2.23235+(0.188018 \times {\rm [Fe/H]})$. This initial data set contained $8535$ stars: $7667$ from SEGUE, $471$ from Legacy and $397$ from MARVELS. Figure 1 shows ${\rm T}_{\rm eff}$ versus ${\rm log(g)}$ for all $8535$ stars, and Figure 2 is a histogram of the metallicities. In Figure 2, the heavy solid histogram shows the metallicity distribution for the full data set, with the histogram for just the SEGUE stars shown as a dotted line, the histogram for the Legacy stars drawn with a light solid line, and the histogram for the MARVELS stars shown with a dashed line. 

\begin{figure}
\includegraphics[width=84mm]{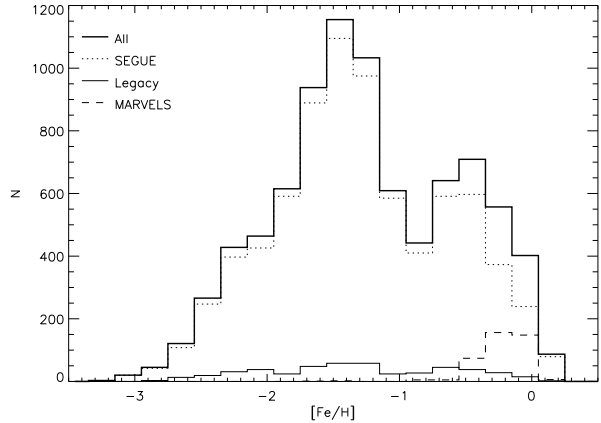}
\caption{Metallicity histogram for our initial data set (heavy solid
line), with histograms for the three data sources overplotted with a
dotted line (SEGUE), light solid line (Legacy) and dashed line (MARVELS).}
\end{figure}

\section{Low-Resolution Analysis}
In order to identify stars with potentially high lithium abundance based on low-resolution spectra, we designed a spectral index, $S(Li)$, that measures the magnitude difference between the integrated flux in the $6708\hbox{\AA}$ resonance line and the integrated flux in a small region of nearby spectrum. It is defined as
\[ 
S(Li) = -2.5 \times \log \frac{\int_{6706}^{6713} I_{\lambda}d\lambda}{\int_{6730}^{6738} I_{\lambda}d\lambda}.
\]
Typically spectral indices are defined for use in small regions of parameter space, and use one or more nearby region(s) of spectrum independent of the property that drives variations in the feature of interest as a comparison band; however, our initial data set contains stars across such a wide range of ${\rm T}_{\rm eff}$, ${\rm log(g)}$ and ${\rm [Fe/H]}$ that there are no nearby regions of spectrum that are the same for all of our stars. This fact introduces some scatter to our $S(Li)$ measurements, requiring extra care in selecting likely Li-enhanced stars for followup observations. 

\begin{figure}
\includegraphics[width=84mm]{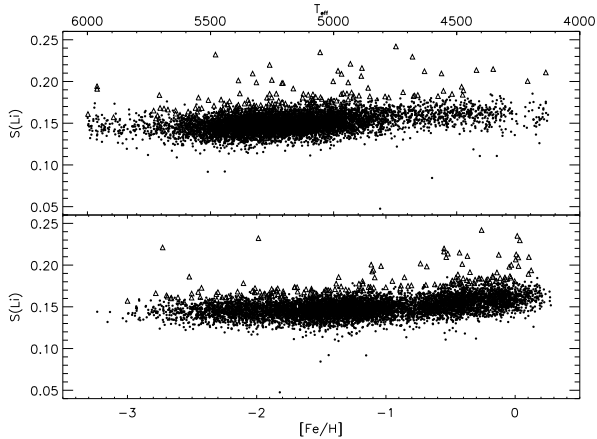}
\caption{Lithium index $S(Li)$ versus ${\rm T}_{\rm eff}$ for the
initial data set (upper panel); $S(Li)$ versus ${\rm [Fe/H]}$ (lower panel). In both panels, stars selected for the secondary data set are shown as open triangles.}
\end{figure}

The upper panel of Figure 3 shows $S(Li)$ versus ${\rm T}_{\rm eff}$ for the initial data set, and the lower panel shows $S(Li)$ versus ${\rm [Fe/H]}$ for the same stars. In each plane, $S(Li)$ has a broad distribution at a fixed value of the independent parameter (${\rm T}_{\rm eff}$ or ${\rm [Fe/H]}$) and a slight dependence on that parameter. To identify outliers in the $S(Li)$-${\rm T}_{\rm eff}$ and $S(Li)$-${\rm [Fe/H]}$ distributions, we select all stars with $S(Li) \geq 0.295 - \frac{{\rm T}_{\rm eff}}{4 \times 10^{4}}$ and $S(Li) \geq 0.18 + \frac{{\rm [Fe/H]}}{100}$ as our secondary data set. There are 162 stars meeting both criteria, and these are plotted as open triangles in Figure 3, while all other stars from the initial data set are represented by small filled circles.

\begin{figure}
\includegraphics[width=84mm]{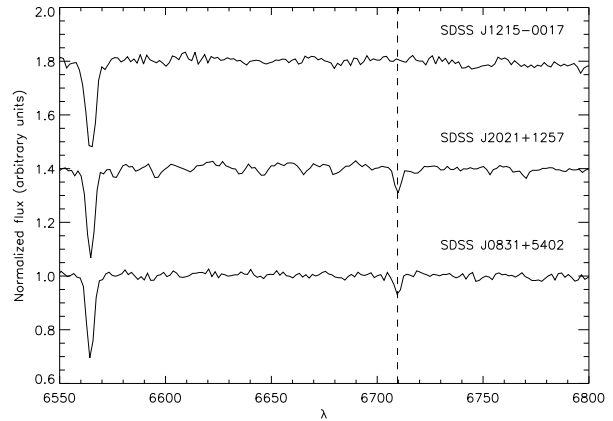}
\caption{Typical spectra from our initial SDSS low-resolution data
set. The lower two spectra appear to have strong absorption in the
$6708\hbox{\AA}$ Li resonance line, while the top spectrum does not.}
\end{figure}

From by-eye examination each of the spectra in this secondary data set of $162$ stars, we found that our selection criteria were generous: fewer than half of the stars selected based on $S(Li)$ - ${\rm T}_{\rm eff}$ - ${\rm [Fe/H]}$ position had $6708\hbox{\AA}$ Li lines that appeared legitimately strong in the SDSS spectra. Figure 4 shows spectra for two of the stars we selected for high-resolution followup (the lower two spectra), as well as one star with little apparent absorption in the $6708\hbox{\AA}$ Li line (top spectrum). We generated low-resolution synthetic spectra based on SSPP stellar parameters to roughly estimate lithium abundance for those potentially lithium-enhanced stars, and selected 36 of them for high-resolution followup observations, with the goal of determining more precise stellar parameters and $\epsilon$(Li) abundances. Table 1 lists SDSS ID number, data source (SEGUE, Legacy or MARVELS), plug-plate number, modified Julian date of observation, fiber ID number, right ascension, declination, g magnitude (V magnitude for the MARVELS stars) and $S(Li)$ measurement for the 36 stars initially flagged for followup spectroscopy. MARVELS stars are too bright to be included in SEGUE photometry, so their apparent $V$ magnitudes were taken from the Vizier catalog-search service \footnote{http://vizier.cfa.harvard.edu/viz-bin/VizieR}. The plate, MJD and fiberID information is given because it is a simple way for finding individual stars in the CAS. 

\section{High-Resolution Analysis}
Followup spectra were obtained for 34 of these 36 stars 
using the High-Resolution Spectrograph (HRS) on the Hobby-Eberly Telescope (HET) at McDonald Observatory \citep{T98}.  
All but one of these 34 stars were first observed as filler time (P4) during 
normal queue observing \citep{SC07} during periods 2010T2 and 2011T1.
This filler time allowed us to 
take an initial short snapshot spectrum to confirm the stength of the Li 
line; 7 were found to not be Li rich and 
27 were confirmed Li rich.   For 23 of these Li rich stars we were able to 
follow-up with additional spectra to get the SNR sufficient for determination of stellar parameters and abundances.  One of the 36 stars (SDSS J0831+5402) had already been abserved as part of the Chemical Abundances of Stars in the Halo (CASH) project \citep{FA08}. 

\begin{table*}
\centering
\caption{Star identifiers, position, photometry and lithium index for the 36 stars selected for followup spectroscopy. Apparent V magnitudes are listed for MARVELS stars.}
\begin{tabular}{l l r r r r r c c}
\hline
SDSS ID & Source & Plate & MJD & FiberID & $\alpha$ & $\delta$ & $g_{0}$ & $S(Li)$\\
\hline
\multicolumn{9}{c}{Confirmed Li-rich with HET}\\
SDSS J2206+4531 & SEGUE & 2556 & 54000 & 640 & 22:06:19.75 & 45:31:57.19 & 15.732 & 0.209\\
SDSS J2353+5728 & SEGUE & 2377 & 53991 & 337 & 23:53:29.76 & 57:28:18.41 & 12.960 & 0.209\\
SDSS J0808-0815 & SEGUE & 2806 & 54425 & 256 & 08:08:34.33 & -08:15:47.30 & 14.598 & 0.198\\
SDSS J0301+7159 & MARVELS & 2843 & 54453 & 634 & 03:01:33.77 & 71:59:11.16 & 10.570 & 0.197\\
SDSS J2019+6012 & SEGUE & 2554 & 54263 & 600 & 20:19:32.44 & 60:12:46.32 & 15.270 & 0.212\\
SDSS J0652+4052 & MARVELS & 2847 & 54452 & 557 & 06:52:37.80 & 40:52:01.44 & 11.170 & 0.212\\
SDSS J0245+7102 & MARVELS & 2843 & 54453 & 168 & 02:45:20.83 & 71:02:02.63 & 11.760 & 0.184\\
SDSS J0632+2604 & SEGUE & 2678 & 54173 & 188 & 06:32:30.90 & 26:04:37.46 & 15.152 & 0.234\\
SDSS J2356+5633 & SEGUE & 2377 & 53991 & 200 & 23:56:52.22 & 56:33:38.83 & 13.080 & 0.210\\
SDSS J0535+0514 & MARVELS & 2841 & 54451 & 299 & 05:35:43.75 & 05:14:22.62 & 12.310 & 0.191\\
SDSS J2200+4559 & SEGUE & 2556 & 54000 & 540 & 22:00:51.22 & 45:59:40.33 & 15.453 & 0.229\\
SDSS J0304+3823 & SEGUE & 2441 & 54065 & 453 & 03:04:37.40 & 38:23:46.12 & 15.357 & 0.181\\
SDSS J1901+3808 & SEGUE & 2536 & 53883 & 306 & 19:01:49.00 & 38:08:51.00 & 14.29 & 0.214\\
SDSS J0720+3036 & SEGUE & 2677 & 54180 & 255 & 07:20:42.79 & 30:36:51.49 & 16.667 & 0.209\\
SDSS J1909+3837 & SEGUE & 2536 & 53883 & 110 & 19:09:29.25 & 38:37:46.23 & 15.764 & 0.214\\
SDSS J1105+2850 & SEGUE & 2870 & 54534 & 599 & 11:05:36.91 & 28:50:08.09 & 17.006 & 0.185\\
SDSS J0654+4200 & MARVELS & 2847 & 54552 & 614 & 06:54:58.00 & 42:00:27.00 & 12.51 & 0.220\\
SDSS J1607+0447 & SEGUE & 2178 & 54629 & 378 & 16:07:09.23 & 04:47:12.73 & 16.139 & 0.173\\
SDSS J1310-0012 & SEGUE & 2901 & 54652 & 191 & 13:10:37.22 & -00:12:44.40 & 15.252 & 0.185\\
SDSS J0936+2935 & SEGUE & 2889 & 54530 & 164 & 09:36:27.44 & 29:35:35.80 & 16.240 & 0.185\\
SDSS J1522+0655 & SEGUE & 2902 & 54629 & 33 & 15:22:11.86 & 06:55:55.22 & 15.912 & 0.165\\
SDSS J0831+5402 & SEGUE & 2316 & 53757 & 395 & 08:31:55.28 & 54:02:45.55 & 15.430 & 0.186\\
SDSS J1432+0814 & SEGUE & 2933 & 54617 & 107 & 14:32:07.15 & 08:14:06.15 & 17.047 & 0.221\\
\hline
\multicolumn{9}{c}{Confirmed Li-rich with HET but not enough SNR for full analysis}\\
SDSS J0515+1558 & SEGUE & 2668 & 54084 & 214 & 05:15:23.17 & 15:58:55.38 & 14.555 & 0.216\\
SDSS J2048+5603 & SEGUE & 2555 & 54265 & 58 & 20:48:46.82 & 56:03:53.65 & 13.423 & 0.200\\
SDSS J2119-0734 & SEGUE & 2305 & 54414 & 230 & 21:19:47.65 & -07:34:26.89 & 14.526 & 0.167\\
SDSS J1214+5027 & SEGUE & 2919 & 54537 & 630 & 12:14:54.88 & 50:27:00.99 & 18.205 & 0.177\\
\hline
\multicolumn{9}{c}{Confirmed not Li-rich with HET}\\
SDSS J1421+0136 & Legacy & 534 & 51997 & 162 & 14:21:28.26 & 01:36:28.35 & 17.077 & 0.200\\
SDSS J1213-0147 & Legacy & 333 & 52313 & 331 & 12:13:53.32 & -01:47:14.93 & 16.558 & 0.180\\
SDSS J1227-0026 & SEGUE & 2558 & 54140 & 115 & 12:27:12.72 & -00:26:58.81 & 16.990 & 0.177\\
SDSS J0144-1005 & SEGUE & 2850 & 54461 & 12 & 01:44:49.17 & -10:05:31.09 & 16.738 & 0.171\\
SDSS J0352-0540 & SEGUE & 2051 & 53738 & 357 & 03:52:43.59 & -05:40:03.32 & 16.487 & 0.168\\
SDSS J1725+0729 & SEGUE & 2797 & 54616 & 304 & 17:25:04.06 & 07:29:45.03 & 14.807 & 0.168\\
SDSS J1849+1931 & SEGUE & 2812 & 54633 & 393 & 18:49:52.58 & 19:31:04.20 & 14.980 & 0.168\\
\hline
\multicolumn{9}{c}{Not observed with HET}\\
SDSS J1824+7911 & SEGUE & 2802 & 54326 & 372 & 18:24:22.51 & 79:11:24.47 & 16.548 & 0.201\\
SDSS J2021+1257 & SEGUE & 2248 & 53558 & 291 & 20:21:25.73 & 12:57:51.33 & 15.263 & 0.198\\

\hline
\end{tabular}
\end{table*}

\subsection{Spectroscopic Observations and Reduction}

HRS was configured to
HRS\_15k\_central\_316g5936\_2as\_0sky\_IS0\_GC0\_2x5 during our HET observations to achieve R=18,000 spectra covering 4233\AA\ to 5900\AA\ on the blue
chip and 6000\AA\ to 7850\AA\ on the red chip. Exposure times per visit ranged from 300 seconds
to 1800 seconds based on the observing conditions.  
Because of the nature of the HET and the generally poor conditions 
of filler time, some targets had multiple visits
taken over a number of days to build up enough S/N to cleanly 
detect the Li feature. 

The spectra were reduced with IRAF\footnote{IRAF (Image Reduction and
Analysis Facility) is distributed by the National Optical Astronomy
Observatories, which are operated by the Association of Universities
for Research in Astronomy, Inc., under contract with the National
Science Foundation.} ECHELLE scripts. The standard IRAF scripts for
overscan removal, bias subtraction, flat fielding and scattered light
removal were employed. For the HRS flat field we masked out
the Li I and Na D regions because the HET HRS flat field lamp suffered from intermittent faint emission lines which would otherwise compromise our analysis. Equivalent widths (EW) were measured
using the IRAF task {\it splot} and forcing the Gaussian fit to
have fixed FWHM for those lines below 150 m\AA\ while allowing the
strong lines to have a variable FWHM. We also fit multiple
Gaussian profiles to any nearby lines to take blends into account. Metallicity and $\alpha$-element abundances were determined through equivalent width analysis; oscillator strengths adopted for the EW analysis were taken from 
\citet{FM06}, \citet{FM07}, \citet{JP10}, and \citet{KM08}. Lithium abundance was determined by spectral synthesis of the 6708\AA\ and 6104\AA\ resonance lines. We also determined carbon abundance and $^{\rm 12}{\rm C}/^{\rm 13}{\rm C}$ ratios, where possible, by synthesis of the 4297\AA\ CH feature. The CH linelist for the carbon analysis was
taken from Plez (priv. comm.); all other lines were taken from a version of the \citet{KB95} linelist modified to fit the strongest lines to the Solar spectrum. Model atmospheres were taken from the Osmarcs 2005
grid \citep{GE08}, interpolating between the grid points. These spherical models
cover metallicity ranges from +1.0 to -3.0 with $[\alpha/Fe]=+0.4$
for $[Fe/H]<=-1.5$, $[\alpha/Fe]=+0.3$ for $[Fe/H]=-0.75$, $[\alpha/Fe]=+0.2$ for $[Fe/H]=-0.5$, $[\alpha/Fe]=+0.1$ for $[Fe/H]=-0.25$ and
$[\alpha/Fe]=+0$. for $[Fe/H]>=0.0$).

The abundance calculations were performed
using a modified version of the 2010 version of the LTE line analysis 
and spectrum synthesis code MOOG \citep{S73}. This code has some significant differences
from the earlier versions including using the predicted abundance vs.
reduced EW for determining the microturbulence. Initial estimates of the stellar parameters were taken from the SSPP pipeline. From this starting point the temperature was adjusted until the there was
no significant slope in the Fe I abundances with respect to line excitation 
potential. The stellar gravity was adjusted such that the Fe I and Fe II 
lines produced the same abundance within the errors. Microturbulence 
was determined by requiring that there was no significant slope in the 
Fe I abundance vs. predicted Fe I reduced equivalent widths based on the 
mean abundance of the lines.

Table 2 lists SDSS ID, along with ${\rm T}_{\rm eff}$, ${\rm log(g)}$ and ${\rm [Fe/H]}$ and their associated errors, plus microturbulent velocity and rotational velocity, where measurable, derived from our high-resolution spectra, and Table 3 lists SDSS ID and abundances of ${\rm [}\alpha{\rm /Fe]}$, ${\rm [C/Fe]}$, $^{12}{\rm C/}^{13}{\rm C}$, and ${\rm log} \epsilon{\rm (Li)}$ determined from synthesis of the $6707 \hbox{\AA}$ and $6104 \hbox{\AA}$ lines. Alpha-element abundances are the error-weighted average of [Mg/Fe], [Ca/Fe] and [Ti/Fe], and the total error on [$\alpha$/Fe] is the larger of the propagated errors or the dispersion in the three input abundances.

\begin{figure}
\includegraphics[width=84mm]{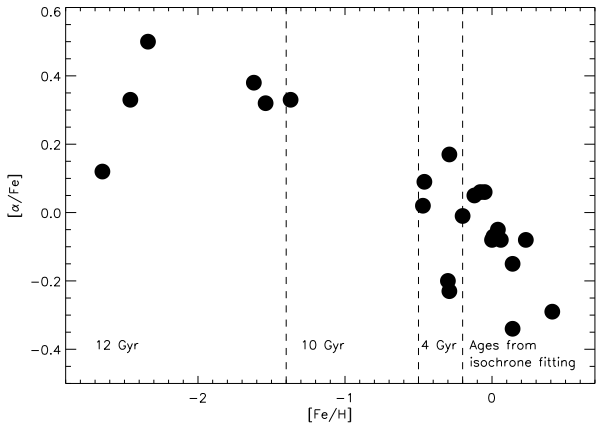}
\caption{The 23 stars for which we derive high-resolution stellar parameters and abundances, in the $\alpha$ -- Fe plane. Dashed lines mark the borders of our rough age estimations.}
\end{figure}

\begin{figure}
\includegraphics[width=84mm]{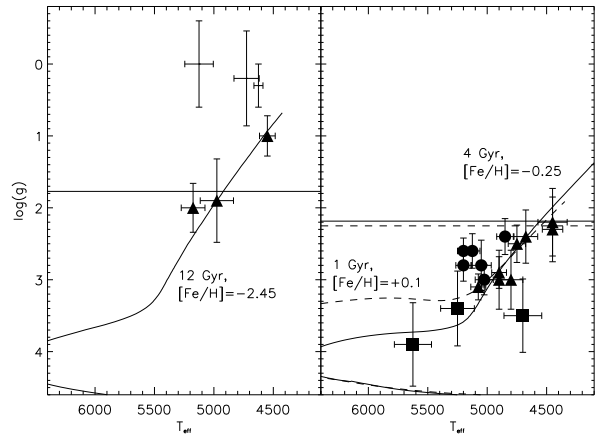}
\caption{Effective temperature and surface gravity from high-resolution analysis, with measurement errors, for our 23 Li-rich stars. Lower-metallicity stars are in the left panel, and higher-metallicity stars are in the right panel. The different symbols represent groups in evolutionary phase, and representative isochrones are overplotted. Horizontal lines mark the position of the RGB bump on each isochrone, as calculated from our empirical relation given in Section 2.2. Triangles are RGB stars, circles are red clump stars, and squares are stars that have not yet reached first dredge-up (whose Li abundances are not relevant to the question of Li production on the RGB). The three plain crosses at the top of the left panel are stars whose log(g) and ${\rm T}_{eff}$ are not compatible with any reasonable isochrone, and we believe them to be post-AGB stars.}
\end{figure}

\subsection{Evolutionary states of the stars}
Previous searches for Li-rich stars have focused on groups of stars with similarities: for example, \citet{LU12} selected only RGB stars in the Galactic bulge, with metallicities in a narrow range, and \citet{KR11} use the {\it Hipparcos} catalog, in which the distances to all stars are known quite precisely, as their initial data set. However, our data set is heterogeneous in terms of age, composition and position in the Galaxy. Based on the metallicities and alpha-enhancements derived from the high-resolution spectra, we divide our Li-rich stars into four broad groups and assign representative ages. Figure 5 shows the four groups in the $\rm{[}\alpha\rm{/Fe]}$ -- $\rm{[Fe/H]}$ plane, which is commonly used to compare the star-formation and chemical-enrichment histories of composite stellar populations in dwarf galaxies to the components of the Milky Way (e.g., \citealt{VT03}; \citealt{SV03}; \citealt{KG08}; \citealt{KC11}). Briefly, moderate- and low-metallicity stars are assumed to be old, while higher-metallicity, non-$\alpha$-enhanced stars are significantly younger. For the highest-metallicity stars in the sample ($\rm{[Fe/H]} \geq -0.2$), we use isochrones to estimate the ages.

Figure 6 demonstrates that our initial ${\rm [\alpha/Fe]}$ -- [Fe/H] age estimation was reasonable:  both panels compare our values of ${\rm T}_{\rm eff}$ and ${\rm log(g)}$ measured from high-resolution spectra against representative isochrones, with older stars in the left panel and younger stars in the right panel, and the majority of stars lie near the giant branch corresponding to the appropriate initial values. The horizontal lines in each panel mark the ${\rm log(g)}$ at the RGB bump for each isochrone shown, calculated from our empirical calibration given in Section 2.2. The squares represent stars that have not yet begun to ascend the RGB, and their high Li abundances are presumably primordial. The triangles represent RGB stars, both young and old, and the circles show red clump stars, which all sit to the left of the isochrone at their metallicity, and are all at intermediate age. The three crosses at the top of the left panel are three low-metallicity stars with parameters that are apparently inconsistent with the isochrones; we will address these stars in Sec. 5. The last two columns of Table 2 give our final estimates for age and evolutionary phase. These values are not given for the three stars with parameters incompatible with the isochrones.

We are able to determine $^{\rm 12}{\rm C}/^{\rm 13}{\rm C}$ for all six of our red clump stars, using synthesis of the 4297\AA\ CH feature. In all of these stars, the ratio is fairly low ($^{\rm 12}{\rm C}/^{\rm 13}{\rm C} \leq 15$), indicating that first dredge-up has efficiently mixed the surface material with the stellar interior.

The Li-rich stars are mainly found low on the RGB, which might be expected from Li production associated with the onset of deep mixing. However, the presence of some bright Li-rich giants and red clump stars indicates that it must also be possible for a star to produce Li during other evolutionary phases. Current models of thermohaline mixing on the red giant branch are fairly unconstrained in terms of whether matter is transported away from the hydrogen-burning shell quickly enough for the Cameron-Fowler machanism to produce Li. Figures 7 and 8 show our resulting Li abundances versus ${\rm log(g)}$ and ${\rm [Fe/H]}$, respectively, using the same symbols as in Figure 6.

\begin{table*}
\centering
\caption{ID and stellar parameters from high-resolution analysis, for the 23 stars for which we could determine high-resolution stellar parameters}
\begin{tabular}{l r r r r r r r r r r r}
\hline
SDSS ID & ${\rm T}_{\rm eff}$ & $\sigma$ & ${\rm log(g)}$ & $\sigma$ & ${\rm [Fe/H]}$ & $\sigma$ & ${\rm v}_{\rm turb} ({\rm km s}^{-1})$ & $\sigma$ & $\rm{v}_{\rm{rot}} ({\rm km s}^{-1})$ & Age (Gyr) & EP\\
\hline
SDSS J2206+4531 & $4700$ & $159$ & $3.5$ & $0.51$ & $+0.41$ & $0.17$ & $1.95$ & $0.29$ & ... & 9 & SGB\\
SDSS J2353+5728 & $5025$ & $76$ & $3.0$ & $0.21$ & $+0.23$ & $0.08$ & $1.67$ & $0.11$ & ... & 4 & RC\\
SDSS J0808-0815 & $4900$ & $103$ & $3.0$ & $0.41$ & $+0.14$ & $0.10$ & $1.72$ & $0.14$ & ... & 2 & RGB\\
SDSS J0301+7159 & $5075$ & $63$ & $3.1$ & $0.18$ & $+0.14$ & $0.06$ & $1.79$ & $0.08$ & ... & 1 & RC\\
SDSS J2019+6012 & $4800$ & $100$ & $3.0$ & $0.41$ & $+0.06$ & $0.10$ & $1.70$ & $0.13$ & ... & 3 & RGB\\
SDSS J0652+4052 & $4900$ & $62$ & $2.9$ & $0.22$ & $+0.04$ & $0.07$ & $1.73$ & $0.08$ & ... & 2 & RC\\
SDSS J0245+7102 & $4850$ & $70$ & $2.4$ & $0.25$ & $+0.01$ & $0.07$ & $1.77$ & $0.07$ & ... & 4 & RC\\
SDSS J0632+2604 & $5050$ & $84$ & $2.8$ & $0.35$ & $+0.00$ & $0.09$ & $1.84$ & $0.11$ & ... & 4 & RC\\
SDSS J2356+5633 & $4750$ & $73$ & $2.5$ & $0.26$ & $-0.05$ & $0.07$ & $2.00$ & $0.07$ & ... & 1 & RGB\\
SDSS J0535+0514 & $5200$ & $64$ & $2.8$ & $0.22$ & $-0.08$ & $0.06$ & $1.66$ & $0.08$ & ... & 4 & RC\\
SDSS J2200+4559 & $4675$ & $100$ & $2.4$ & $0.37$ & $-0.12$ & $0.11$ & $1.99$ & $0.39$ & $6$ & 2 & RGB\\
SDSS J0304+3823 & $5125$ & $64$ & $2.6$ & $0.24$ & $-0.20$ & $0.07$ & $1.33$ & $0.09$ & ... & 4 & RC\\
SDSS J1901+3808 & $4450$ & $86$ & $2.3$ & $0.45$ & $-0.29$ & $0.08$ & $1.78$ & $0.09$ & ... & 4 & RGB\\
SDSS J0720+3036 & $5250$ & $142$ & $3.4$ & $0.52$ & $-0.29$ & $0.15$ & $1.40$ & $0.25$ & $10$ & 4 & SGB\\
SDSS J1909+3837 & $4450$ & $123$ & $2.2$ & $0.47$ & $-0.30$ & $0.13$ & $2.13$ & $0.13$ & ... & 4 & RGB\\
SDSS J1105+2850 & $5625$ & $156$ & $3.9$ & $0.58$ & $-0.46$ & $0.18$ & $1.05$ & $0.31$ & ... & 4 & SGB\\
SDSS J0654+4200 & $5200$ & $50$ & $2.6$ & $0.18$ & $-0.47$ & $0.05$ & $1.88$ & $0.06$ & $12.5$ & 4 & RC\\
SDSS J1607+0447 & $5175$ & $100$ & $2.0$ & $0.34$ & $-1.37$ & $0.13$ &$1.49$ & $0.15$ & ... & 10 & RGB\\
SDSS J1310-0012 & $4550$ & $66$ & $1.0$ & $0.28$ & $-1.54$ & $0.06$ & $1.74$ & $0.07$ & ... & 12 & RGB\\
SDSS J0936+2935 & $5125$ & $120$ & $0.0$ & $0.60$ & $-1.62$ & $0.15$ & $2.59$ & $0.23$ & ... & ... & ... \\
SDSS J1522+0655 & $4975$ & $140$ & $1.9$ & $0.58$ & $-2.34$ & $0.15$ & $1.89$ & $0.18$ & ... & 12 & RGB\\
SDSS J0831+5402 & $4725$ & $106$ & $0.2$ & $0.66$ & $-2.46$ & $0.12$ & $2.32$ & $0.23$ & ... & ... & ... \\
SDSS J1432+0814 & $4625$ & $38$ & $0.3$ & $0.3$ & $-2.65$ & $0.06$ & $1.90$ & $0.13$ & $9.1$ & ... & ... \\
\hline
\end{tabular}
\end{table*}

\begin{table*}
\centering
\caption{ID and abundances from high-resolution analysis, for the 23 stars for which we could determine high-resolution stellar parameters}
\begin{tabular}{l r r r r r r r r r}
\hline
SDSS ID & ${\rm [\alpha/Fe]}$ & $\sigma$ & ${\rm [C/Fe]}$ & $\sigma$ & $^{\rm 12}{\rm C}/^{\rm 13}{\rm C}$ & ${\rm log \epsilon (Li)}_{\rm 6708}$ & $\sigma$ & ${\rm log \epsilon (Li)}_{\rm 6104}$ & $\sigma$ \\
\hline
SDSS J2206+4531  & $-0.29$ & $0.07$ & $-0.5$ & $0.3$ & ... & $2.80$ & $0.30$ & $\leq 2.9$ & ... \\
SDSS J2353+5728 & $-0.08$ & $0.06$ & $-0.5$ & $0.2$ & $8$ & $3.10$ & $0.15$ & $\leq 3.2$ & ... \\
SDSS J0808-0815  & $-0.34$ & $0.18$ & $-0.6$ & $0.2$ & ... & $2.45$ & $0.15$ & $\leq 2.7$ & ... \\
SDSS J0301+7159 & $-0.15$ & $0.08$ & $-0.6$ & $0.1$ & $15$ & $2.70$ & $0.10$ & $\leq 2.9$ & ... \\
SDSS J2019+6012  & $-0.08$ & $0.06$ & $-0.7$ & $0.2$ & ... & $2.77$ & $0.20$ & $\leq 2.8$ & ... \\
SDSS J0652+4052 & $-0.05$ & $0.06$ & $-0.4$ & $0.2$ & $15$ & $3.30$ & $0.20$ & $3.0$ & $0.3$ \\
SDSS J0245+7102 & $-0.07$ & $0.04$ & $-0.4$ & $0.1$ & $15$ & $2.58$ & $0.10$ & $2.4$ & $0.4$ \\
SDSS J0632+2604 & $-0.08$ & $0.05$ & $-0.4$ & $0.1$ & ... & $4.20$ & $0.2$ & $4.1$ & $0.2$ \\
SDSS J2356+5633 & $+0.06$ & $0.05$ & $-0.7$ & $0.1$ & ... & $2.88$ & $0.15$ & $2.9$ & $0.2$ \\
SDSS J0535+0514 & $+0.06$ & $0.04$ & $-0.5$ & $0.2$ & $15$ & $2.90$ & $0.10$ & $2.7$ & $0.3$ \\
SDSS J2200+4559  & $+0.05$ & $0.08$ & $-0.8$ & $0.2$ & ... & $3.05$ & $0.20$ & $3.0$ & $0.2$ \\
SDSS J0304+3823 & $-0.01$ & $0.04$ & $-0.3$ & $0.2$ & ... & $2.40$ & $0.10$ & $\leq 2.6$ & ... \\
SDSS J1901+3808 & $+0.17$ & $0.07$ & $-0.7$ & $0.1$ & ... & $2.52$ & $0.15$ & $2.4$ & $0.3$ \\
SDSS J0720+3036  & $-0.23$ & $0.13$ & $-0.4$ & $0.2$ & ... & $4.55$ & $0.20$ & $4.2$ & $0.2$ \\
SDSS J1909+3837  & $-0.20$ & $0.07$ & $\leq -0.4$ & ... & ... & $1.95$ & $0.15$ & $\leq 2.2$ & ... \\
SDSS J1105+2850  & $+0.09$ & $0.48$ & $-0.4$ & $0.4$ & ... & $3.24$ & $0.30$ & $\leq 2.9$ & ... \\
SDSS J0654+4200 & $+0.02$ & $0.06$ & $-0.4$ & $0.1$ & $4$ & $3.30$ & $0.20$ & $3.1$ & $0.3$ \\
SDSS J1607+0447 & $+0.33$ & $0.06$ & $-0.6$ & $0.2$ & ... & $2.55$ & $0.15$ & $2.7$ & $0.3$ \\
SDSS J1310-0012 & $+0.32$ & $0.09$ & $-0.7$ & $0.2$ & ... & $2.15$ & $0.15$ & $2.2$ & $0.3$ \\
SDSS J0936+2935  & $+0.38$ & $0.11$ & $0.3$ & $0.4$ & ... & $2.70$ & $0.30$ & $\leq 3.0$ & ... \\
SDSS J1522+0655  & $+0.50$ & $0.20$ & $-0.7$ & $0.3$ & ... & $2.25$ & $0.20$ & $2.6$ & $0.4$ \\
SDSS J0831+5402  & $+0.33$ & $0.11$ & $0.1$ & $0.3$ & ... & $2.55$ & $0.20$ & $\leq 2.7$ & ... \\
SDSS J1432+0814  & $+0.12$ & $0.10$ & $-0.2$ & $0.2$ & ... & $4.02$ & $0.30$ & $3.6$ & $0.3$ \\

\hline
\end{tabular}
\end{table*}

\section{Discussion}
We have identified 23 new Li-rich post-main-sequence stars: 5 bright red giants, 6 red clump stars, 6 RGB-bump stars, 3 lower-RGB stars and 3 post-AGB stars. Although schematic models exist for the production of Li at onset of deep mixing, near the RGB bump, there is not a clear theoretical explanation for the processes that produce and destroy Li in post-main-sequence stars. 

\begin{figure}
\includegraphics[width=84mm]{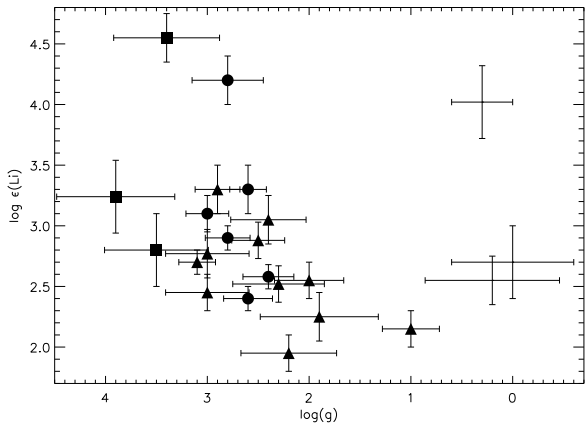}
\caption{Lithium abundance versus surface gravity for the stars in our sample. The typical value for ${\rm log }\epsilon{\rm(Li)}$ decreases in the more evolved stars. Symbols are as in Figure 6, and there is no clear difference in the behavior of RGB stars and red clump stars in this plane.}
\end{figure}

There are no targets in common between our data set and the six recent studies mentioned in the Introduction. However, our initial search of the DR7 CAS included SEGUE observations in Galactic globular and open clusters, allowing a possible overlap with the studies of \citet{P86} and \citet{PS88}. Our star SDSS J2356+5633 is the known Li-rich giant NGC 7789 K301 \citep{P86}. It is chosen as a likely followup candidate according to our low-resolution selection criteria, and our high-resolution analysis confirms its high Li abundance. The fact that this star is in a cluster can be used to check our analysis. The mean metallicity of NGC 7789 is $-0.04 \pm 0.05$ and $0.04 \pm 0.07$ from \citet{TE05} and \citet{PC10}, respectively, which is consistent with our value of $-0.05 \pm 0.07$. In addition, for this star \citet{P86} note that it is a weak G-band star, and we find that this star has one of the lowest carbon abundance ratios in our sample, [C/Fe]$= -0.7$. Further, the position of K301 on the CMD of NGC 7789 suggests that it may be an AGB star, or its low carbon abundance may cause the star to move blueward in the CMD via the Bond-Neff Effect \citep{BN69}.

We do not recover the Li-rich stars discussed in \citet{RF08}, \citet{KR09}, \citet{CF98}, \citet{K99}, \citet{SS99} or \citet{CB00}, which were not observed by SEGUE. Two of the Li-rich stars we selected for high-resolution followup are serendipitously located in the Kepler field. SDSS J1901+3808 is KIC 2968828, which has asteroseismology-based stellar parameters of log(g)$= 2.393$, ${\rm T}_{\rm eff}= 4270$, [Fe/H]$= -0.823$. While the surface gravity is a good match to our determination, the temperature and metallicity are quite different, and we believe that the discrepancy is likely due to uncertainties in the Kepler anaylsis techniques. SDSS J1909+3837 is KIC 3531579, which does not have any publically available Kepler data yet.

The three stars in our high-resolution data set that have stellar parameters incompatible with the isochrones are quite unusual. With their very low surface gravities, we identify them as post-AGB stars. The metallicities we derive for two of them are quite low ([Fe/H]$\sim -2.5$). With these very low metallicities, these two stars are marginally consistent with being RGB-tip stars, but the extremely low gravities are also consistent with a known population of metal-deficient post-AGB stars \citep{P00}. We note that it has been hypothesized that very low metallicities in post-AGB stars are a result of dust formation in extremely cool atmospheres and are not representative of the initial stellar composition. 

\subsection{Internal and external sources of lithium}
The initial photometric association of Li-rich stars with the RGB bump and the red clump led naturally to a suggestion \citep{CB00} that the destruction of the $\mu$-barrier created by first dredge-up causes a brief episode of Li production. This cannot be the explanation for high Li abundances in bright giants, far from the RGB bump, but any process that causes rapid mass motion between the envelope and the hydrogen-burning shell could in principle produce Li at the surface via the Cameron-Fowler mechanism: $^{\rm 3}{\rm He(}\alpha ,\gamma {\rm )}^{\rm 7}{\rm Be(e}^{\rm -}{\rm ,}\nu {\rm )}^{\rm 7}{\rm Li}$.

\begin{figure}
\includegraphics[width=84mm]{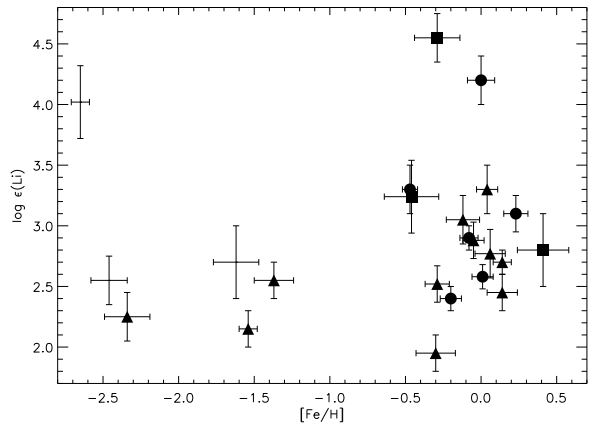}
\caption{Lithium abundance versus metallicity for the stars in our sample. The higher-metallicity stars show a broader range in ${\rm log }\epsilon{\rm(Li)}$, and a higher mean, than the lower-metallicity stars. The rate of deep mixing is a sensitive function of stellar metallicity, so this may indicate that higher-metallicity stars are better able to produce lithium, or less able to destroy it. Symbols are as in Figure 6.}
\end{figure}

One potential mechanism for such stirring is proposed by \citet{D12}, who note that, in their models of red giants with rapid interior rotation, the temperature gradient in the region between the hydrogen-burning shell and the envelope can become quite large, permitting deep mixing to operate much more quickly than usual. One consequence of this accelerated deep mixing is that the photometric loop executed during the RGB bump is much more extended, pushing RGB-bump stars into the same color-magnitude region as red clump stars. This indicates that perhaps the Li-rich stars that have been identified as red clump stars (including the six in this study) are actually RGB-bump stars with rapid internal rotation. Asteroseismic data on these stars could clarify their evolutionary phase. This process is an interesting possibility, but does not explain Li-rich bright RGB stars.

The engulfement of a giant planet has also been suggested \citep{CC12} as a way to ``replenish'' the surface Li abundance of a red giant. This process should be distinguishable from intrinsic Li production by deep mixing, since the $^{\rm 12}{\rm C/}^{\rm 13}{\rm C}$ ratio should rise following planet accretion, but should fall as a result of deep mixing. This is a process that could in principle happen at any point in stellar evolution, though the dramatic radial expansion that occurs as a star ascends the RGB would make it far more likely to happen during that phase.

\section{Summary}
The work presented here demonstrates the potential of large spectroscopic surveys such as SEGUE as a source of unusual targets for followup observations. The ability to select our initial data set from a very broad parameter space allows us to search for Li-rich stars in a relatively unbiased way. Other followup studies based on SEGUE data (e.g., \citealt{CB11}; \citealt{BC11}; \citealt{AB12}) follow the same approach: an initial search in a large, multidimensional parameter space, which introduces some heterogeneity in the data set, but can avoid biases inherent in more focused studies.

Although we agree with \citet{KF12} that Li-rich field giants are perfectly normal stars in most respects, we must disagree with their conclusion, originally expressed by \citet{dDd96}, that the onset of deep mixing produces a brief phase of Li enrichment in all stars. Such a universal process would imply that Li-rich stars should only be found near the RGB bump, which they are not. We propose instead that deep mixing is capable of producing Li enrichment at any point above the RGB bump, based on thermohaline mixing models. This does not explain what triggers a phase of Li production, but does fit comfortably with the the wide range of evolutionary phase that Li-rich stars occupy, and the short duration of the phase, which is inferred from the rarity of these stars.

The combination of the recent large-sample studies of Li-rich stars is extremely important for advancing our understanding of the origins of these stars. Primarily, the new data demonstrates the ubiquity of Li-rich stars across Galactic and extragalactic environments - Li production is clearly a stochastic and short-lived, possibly recurrent event in stellar evolution. The next significant step will need to take an approach similar to the study by \citet{CC12}: identifying unique features of the various proposed sources for Li enrichment in post-main-sequence stars, such as correlated variations in particular elemental or isotopic abundances, and searching specifically for those signals.

\bibliography{li}

\section*{Acknowledgements}
We appreciate thoughtful comments from an anonymous referee, which have made this paper clearer. 

During this work, SLM was supported by Sonderforschungsbereich SFB 881 ``The Milky Way System'' (subprojects A2 and A5) of the German Research Foundation (DFG).

The Hobby-Eberly Telescope (HET) is a joint project of the University of Texas at Austin, the Pennsylvania State University, Ludwig-Maximillians-Universit\"{a}t M\"{u}nchen, and Georg-August-Universit\"{a}t G\"{o}ttingen. The HET is named in honor of its principal benefactors, William P. Hobby and Robert E. Eberly.

Funding for SDSS-III has been provided by the Alfred P. Sloan Foundation, the Participating Institutions, the National Science Foundation, and the U.S. Department of Energy. The SDSS-III web site is http://www.sdss3.org/.

SDSS-III is managed by the Astrophysical Research Consortium for the Participating Institutions of the SDSS-III Collaboration including the University of Arizona, the Brazilian Participation Group, Brookhaven National Laboratory, University of Cambridge, University of Florida, the French Participation Group, the German Participation Group, the Instituto de Astrofisica de Canarias, the Michigan State/Notre Dame/JINA Participation Group, Johns Hopkins University, Lawrence Berkeley National Laboratory, Max Planck Institute for Astrophysics, New Mexico State University, New York University, Ohio State University, Pennsylvania State University, University of Portsmouth, Princeton University, the Spanish Participation Group, University of Tokyo, University of Utah, Vanderbilt University, University of Virginia, University of Washington, and Yale University.

This research has made use of the VizieR catalogue access tool, CDS, Strasbourg, France. The original description of the VizieR service was published in A\&AS 143, 23.

\newpage

\appendix
\section{Comparison of high-resolution and SSPP stellar parameters}
During the course of this project, revised SSPP stellar parameters (including effective temperature, surface gravity and metallicity) were published for our stars as part of SDSS Data Release 9 \citep{DR9}. Since we have derived atmospheric parameters from our high-resolution spectra, we present here a brief study of the comparison between the DR9 SSPP and high-resolution parameters, for the 20 stars with reasonable high-resolution parameters. High-resolution stellar parameters are listed in Table 2; DR9 SSPP parameters for the same stars are given in Table A1.

Figures A1, A2 and A3 compare the high-resolution values for ${\rm T}_{eff}$, ${\rm log(g)}$ and ${\rm [Fe/H]}$, respectively, to the DR9 SSPP values, for the 20 stars with reasonable high-resolution parameters. There is scatter, but not a significant trend, between the high-resolution and SSPP determinations of ${\rm T}_{eff}$ and ${\rm [Fe/H]}$, but we find a significant trend with little scatter between high-resolution and SSPP determinations of ${\rm log(g)}$. We also find a significant trend, with a slope of $102 \pm 39$ K/dex, between ${\rm T}_{\rm eff}^{\rm HRS} - {\rm T}_{\rm eff}^{\rm SSPP}$ and [Fe/H]$^{\rm SSPP}$, indicating that the SSPP algorithms are somewhat susceptible to temperature-metallicity degeneracy. The fact that we use different iron lines to determine [Fe/H] for the metal-poor and metal-rich stars in our sample could also contribute to this trend, but we would not expect that to cause an effect as large as what is observed.

\newpage
\begin{figure}
\includegraphics[width=84mm]{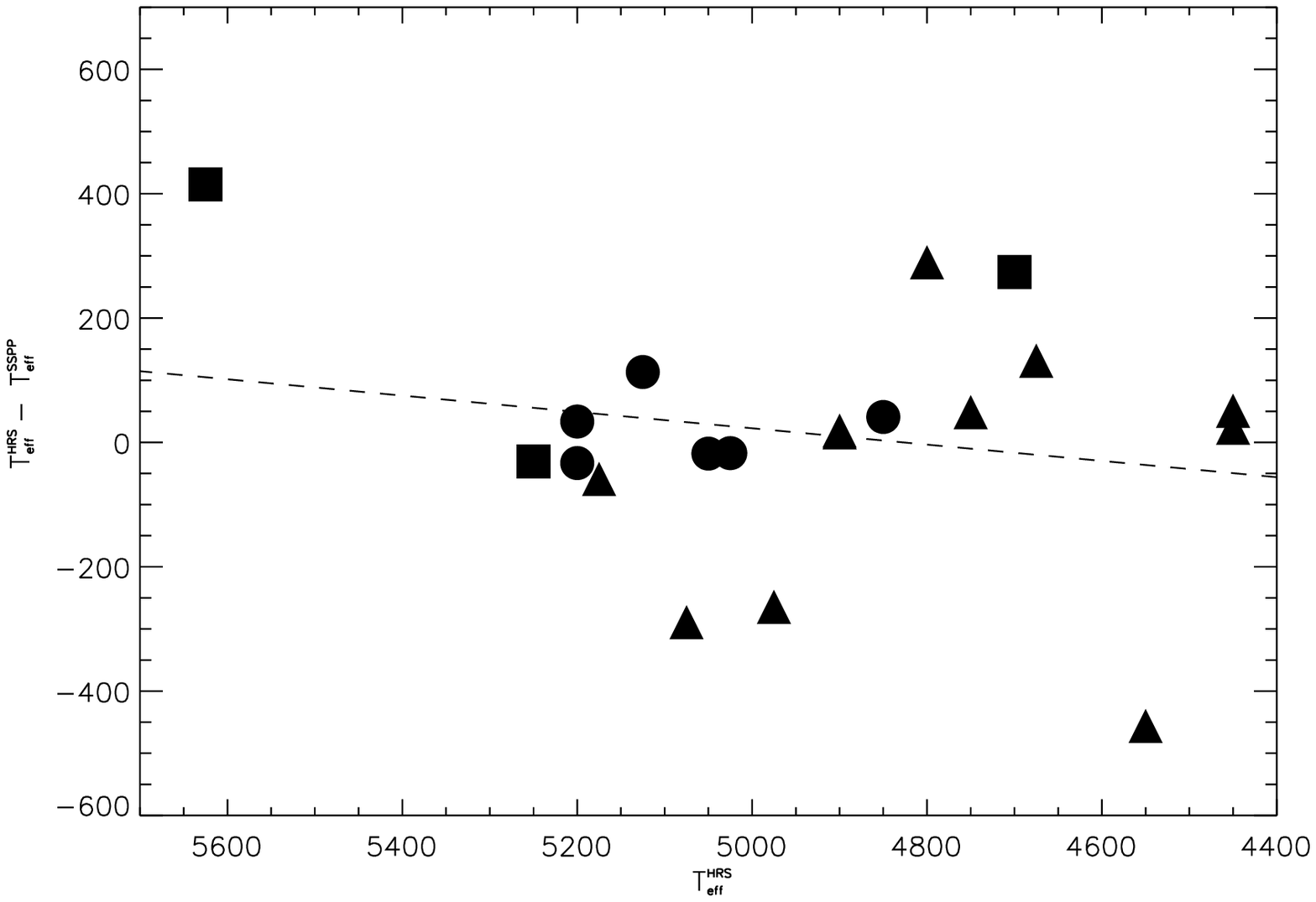}
\caption{Difference between high-resolution and DR9 SSPP determinations of ${\rm T}_{eff}$, as a function of high-resolution ${\rm T}_{eff}$. The best-fit dashed line is consistent with no offset between the two temperature scales, but significant scatter. Symbols are as in Figure 6.}
\end{figure}

\begin{figure}
\includegraphics[width=84mm]{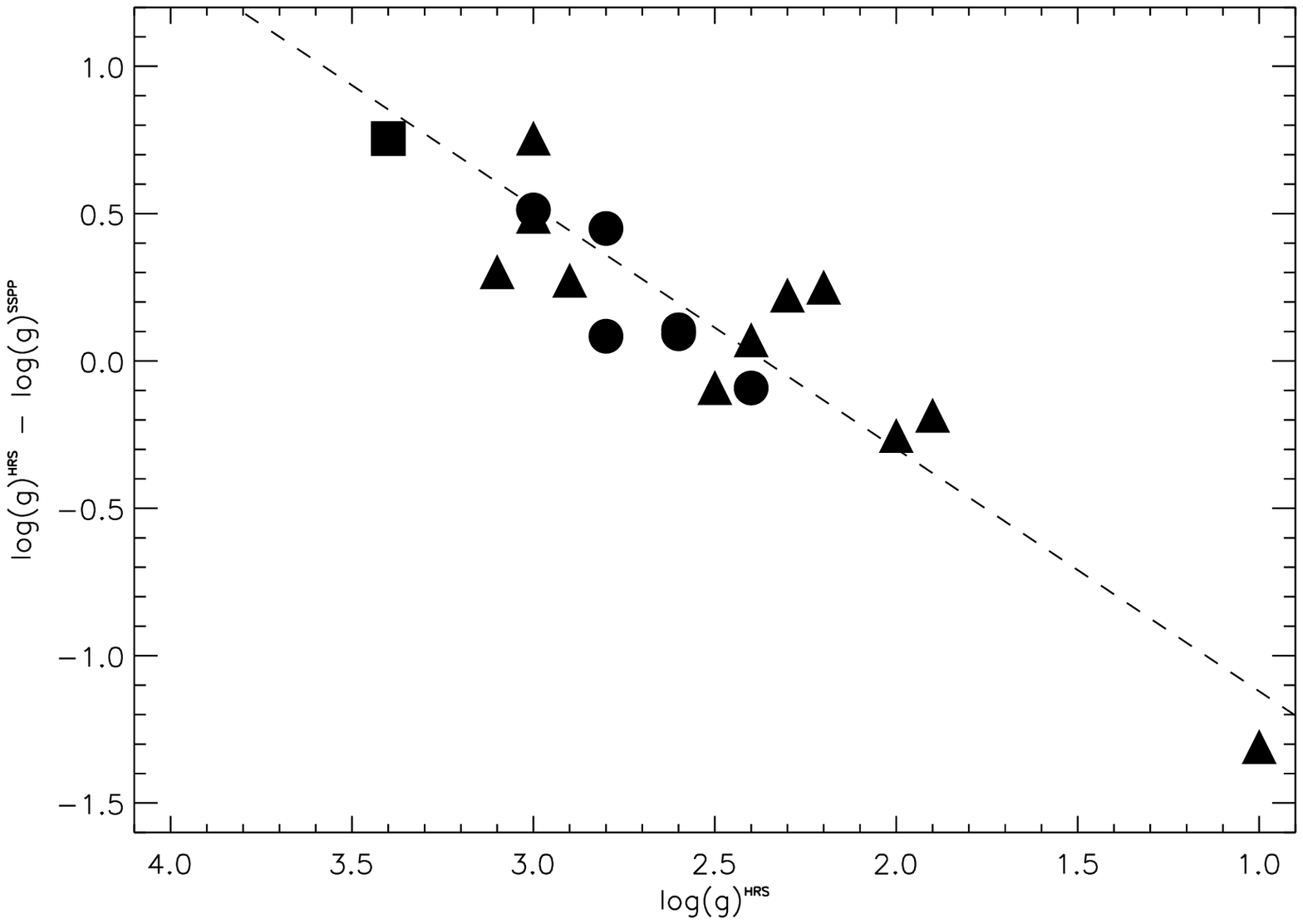}
\caption{Difference between high-resolution and DR9 SSPP determinations of ${log(g)}$, as a function of high-resolution ${\rm log(g)}$. The best-fit dashed line showes a significant trend. The SSPP is known to have difficulty determining ${\rm log(g)}$ in low-gravity stars, since the indicators it uses lose sensitivity; however, the trend to high gravity is unexpected. Symbols are as in Figure 6.}
\end{figure}

\begin{figure}
\includegraphics[width=84mm]{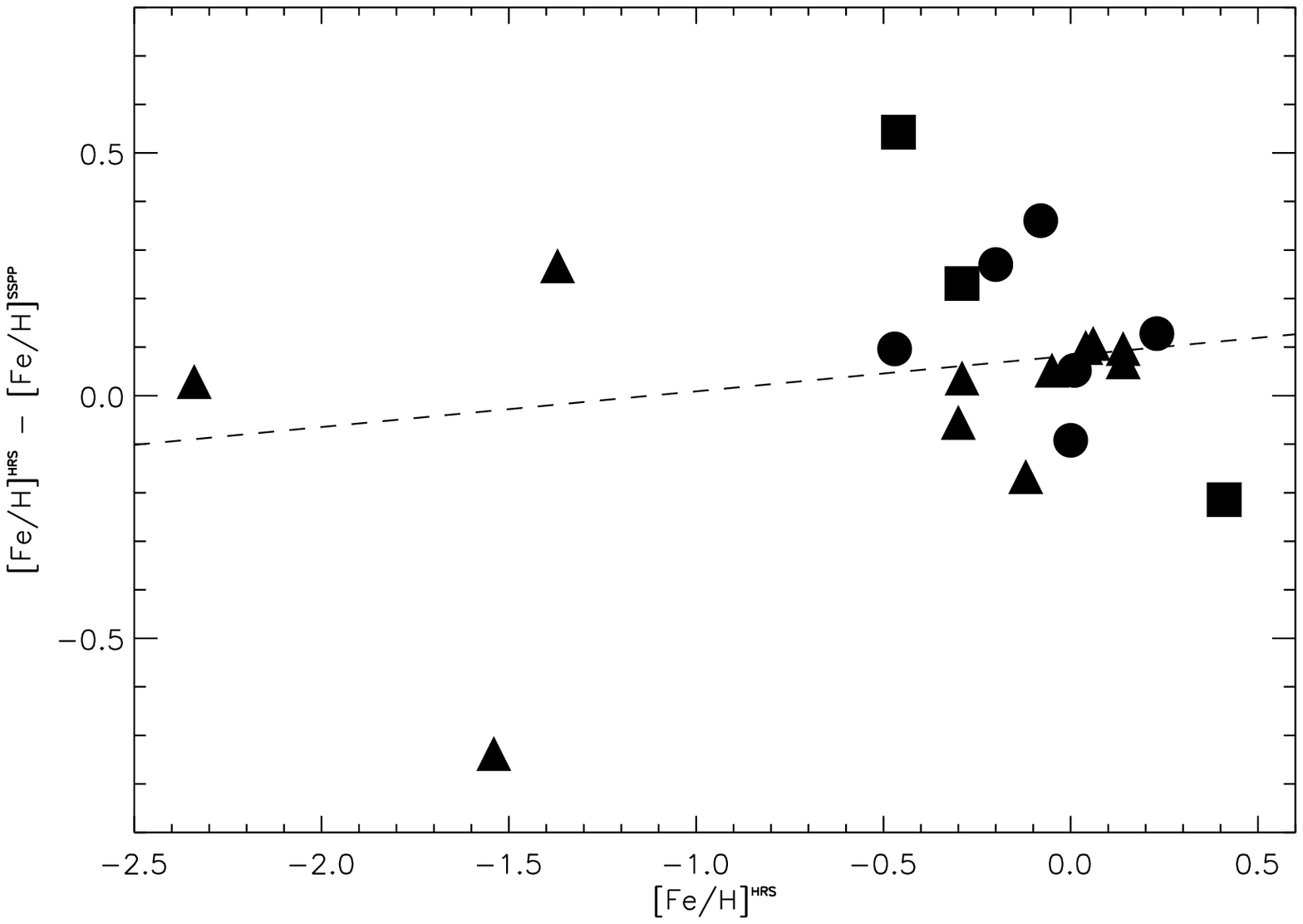}
\caption{Difference between high-resolution and DR9 SSPP determinations of ${\rm [Fe/H]}$, as a function of high-resolution ${\rm [Fe/H]}$. A best-fit line is shown; however, this distribution is consistent with a small offset between the two metallicity scales and $\simeq 0.3$ dex scatter. Symbols are as in Figure 6.}
\end{figure}

\begin{table*}
\centering
\caption{ID and stellar parameters from the DR9 SSPP, for the 23 stars for which we could determine high-resolution stellar parameters}
\begin{tabular}{l r r r r r r}
\hline
SDSS ID & ${\rm T}_{\rm eff}$ & $\sigma$ & ${\rm log(g)}$ & $\sigma$ & ${\rm [Fe/H]}$ & $\sigma$\\
\hline
SDSS J2206+4531 & $4426$ & $15$ & $2.23$ & $0.08$ & $0.62$ & $0.007$\\
SDSS J2353+5728 & $5042$ & $88$ & $2.49$ & $0.09$ & $0.10$ & $0.036$\\
SDSS J0808-0815 & $4884$ & $9$ & $2.51$ & $0.15$ & $0.04$ & $0.009$ \\
SDSS J0301+7159 & $5364$ & $147$ & $2.80$ & $0.14$ & $0.07$ & $0.05$ \\
SDSS J2019+6012 & $4510$ & $70$ & $2.24$ & $0.13$ & $-0.05$ & $0.008$ \\
SDSS J0652+4052 & $4881$ & $88$ & $2.63$ & $0.16$ & $-0.06$ & $0.043$ \\
SDSS J0245+7102 & $4809$ & $113$ & $2.49$ & $0.21$ & $-0.04$ & $0.034$ \\
SDSS J0632+2604 & $5068$ & $25$ & $2.72$ & $0.22$ & $0.09$ & $0.074$ \\
SDSS J2356+5633 & $4702$ & $72$ & $2.59$ & $0.32$ & $-0.10$ & $0.009$ \\
SDSS J0535+0514 & $5166$ & $47$ & $2.35$ & $0.18$ & $-0.44$ & $0.058$ \\
SDSS J2200+4559 & $4544$ & $41$ & $2.33$ & $0.12$ & $0.05$ & $0.003$ \\
SDSS J0304+3823 & $5012$ & $30$ & $2.49$ & $0.14$ & $-0.47$ & $0.031$ \\
SDSS J1901+3808 & $4399$ & $88$ & $2.08$ & $0.08$ & $-0.33$ & $0.014$ \\
SDSS J0720+3036 & $5281$ & $79$ & $2.65$ & $0.10$ & $-0.52$ & $0.036$ \\
SDSS J1909+3837 & $4426$ & $110$ & $1.95$ & $0.31$ & $-0.25$ & $0.073$ \\
SDSS J1105+2850 & $5210$ & $56$ & $2.63$ & $0.10$ & $-1.00$ & $0.026$ \\
SDSS J0654+4200 & $5233$ & $20$ & $2.51$ & $0.06$ & $-0.57$ &$ 0.031$ \\
SDSS J1607+0447 & $5234$ & $56$ & $2.25$ & $0.14$ & $-1.64$ & $0.045$ \\
SDSS J1310-0012 & $5006$ & $244$ & $2.31$ & $0.20$ & $-0.80$ & $0.062$ \\
SDSS J0936+2935 & $5334$ & $35$ & $2.03$ & $0.07$ & $-1.28$ & $0.058$ \\
SDSS J1522+0655 & $5240$ & $46$ & $2.08$ & $0.20$ & $-2.37$ & $0.034$ \\
SDSS J0831+5402 & $4826$ & $80$ & $0.79$ & $0.31$ & $-2.66$ & $0.053$ \\
SDSS J1432+0814 & $4664$ & $239$ & $1.43$ & $0.29$ & $-3.14$ & $0.072$ \\
\hline
\end{tabular}
\end{table*}

\end{document}